# GPU-based fast gamma index calcuation

**Xuejun Gu, Xun Jia, and Steve B. Jiang**

Center for Advanced Radiotherapy Technologies and Department of Radiation Oncology, University of California San Diego, La Jolla, CA 92037-0843

E-mail: sbjiang@ucsd.edu

The $\gamma$-index dose comparison tool has been widely used to compare dose distributions in cancer radiotherapy. The accurate calculation of $\gamma$-index requires an exhaustive search of the closest Euclidean distance in the high-resolution dose-distance space. This is a computational intensive task when dealing with 3D dose distributions. In this work, we combine a geometric method (Ju et al. *Med Phys* **35** 879-87, 2008) with a radial pre-sorting technique (Wendling et al. *Med Phys* **34** 1647-54, 2007), and implement them on computer graphics processing units (GPUs). The developed GPU-based γ-index computational tool is evaluated on eight pairs of IMRT dose distributions. The GPU implementation achieved 20x~30x speedup factor compared to CPU implementation and γ-index calculations can be finished within a few seconds for all 3D testing cases. We further investigated the effect of various factors on both CPU and GPU computation time. The strategy of pre-sorting voxels based on their dose difference values speed up the GPU calculation by about 2-4 times. For $n$-dimensional dose distributions, $\gamma$-index calculation time on CPU is proportional to the summation of $\gamma^n$ over all voxels, while that on GPU is effected by $\gamma^n$ distributions and is approximately proportional to the $\gamma^n$ summation over all voxels. We found increasing dose distributions resolution leads to quadratic increase of computation time on CPU, while less-than-quadratic increase on GPU. The values of dose difference (DD) and distance-to-agreement (DTA) criteria also have their impact on $\gamma$-index calculation time.

## 1. Introduction

The γ-index concept introduced by Low *et al* (Low *et al.* 1998) has been widely used to compare two dose distributions in cancer radiotherapy. The original γ-index calculation algorithm of Low *et al* (Low *et al.* 1998) has been improved for better accuracy and/or efficiency (Depuydt *et al.* 2002; Bakai *et al.* 2003; Stock *et al.* 2005; Jiang *et al.* 2006; Spezi and Lewis 2006). However, since these modified algorithms still involve computational intensive tasks such as interpolation of dose grid and exhaustive search, it is very time-consuming (*e.g.*, many minutes) to compare two 3D dose distributions of clinically relevant sizes.

In more recent years, much effort has been invested to develop fast and/or accurate γ-index calculation algorithms. Wendling *et al.* (Wendling *et al.* 2007) speeded up the exhaustive search by pre-sorting involved evaluation dose points with respect to their spatial distance to reference dose point and performing interpolation on-fly in a fixed searching radius region. This fixed search region induces overestimation of γ-index values at certain case if dose difference values are very large inside the search region and has sharp drop just beyond the search region boundary. This algorithm also relies on fine dose interpolation to secure accuracy. The geometric interpretation of γ-index evaluation technique proposed by Ju *et. al.* (Ju *et al.* 2008) implies a linear interpolation by calculating the distance from a reference point to a subdivided simplex formed by evaluation dose points in search regions. Thus, high accuracy and efficiency can be achieved without interpolating dose grid to fine resolution. However, searching the closest distance over all subdivided simplexes is still time-consuming. Later, Chen *et. al*. (Chen *et al.* 2009) reports a method based on using fast Euclidean distance transform (EDT) of quantized $n$-dimensional dose distributions. Fast γ-index evaluation can be achieved with complexity of $O(N^n M)$, where $N$ is the size of dose distribution in each dimension, $n$ is the number of dimension, and $M$ is the number of quantized values for dose distribution. This method brings in discretization errors when quantizing dose distributions. It also requires $M$ time's more memory space of original searching based algorithm. Thus, a full 3D application of EDT method is limited by its memory requirement. The searching based algorithm's complexity is $O(N^n N_s)$, where $N_s$ represents exhaustive search steps For cases where γ is not very large, where $N_s$ is much smaller than $M$, this EDT method loses its advantage of efficiency. Recently, Yuan *et. al.* (Yuan and Chen 2010) proposes a technique using a *k*-d tree technique for nearest neighbor searching. The searching time for $N^n$ voxels dose distribution can be reduced to $(N^n)^{1/k}$, where $2 < k < 3$ for 2D and 3D dose distributions. However, this method requires interpolating dose grid to secure accuracy. Moreover, in certain cases, the overhead of *k*-d tree construction time is longer than γ-index calculation time.

Another approach to speed up γ-index evaluation is to implement an accurate algorithm on graphics processing unit (GPU) platform. GPUs have recently been introduced into the radiotherapy community to accelerate computational tasks including CBCT reconstruction, deformable image registration, dose calculation, and treatment plan optimization (Jia *et al.* 2010b; Samant *et al.* 2008; Sharp *et al.* 2007; Jacques *et al.*

2008; Hissoiny *et al.* 2009; Gu *et al.* 2010; Gu *et al.* 2009; Jia *et al.* 2010a; Men *et al.* 2009; Men *et al.* 2010a; Men *et al.* 2010b). GPU are especially well-suited for problems that can be expressed as data-parallel computations (NVIDIA 2010). γ-index calculation belongs to this category, because the evaluation of each reference point is totally independent. Instead of implementing the memory demanding EDT method or the large overhead *k*-d tree method, we decide to combine the accuracy of geometric interpretation technique (Ju *et al.* 2008) and the efficiency of the pre-sorting technique (Wendling *et al.* 2007). We will also revise this modified algorithm to make it GPU-friendly and then implement it on GPU to achieve both accuracy and high efficiency.

The remainder of this paper is organized as follows. In Section 2, we will discuss the modified γ-index algorithm and its implementation on GPU. Section 3 will present the evaluation of our GPU-based algorithm using eight 3D IMRT dose distributions pairs. We will first study the speedup factor achieved by GPU implementation. We will further study the effects of dose difference sorting and the $r$-index values on the computation time. We will also investigate how the dose distribution resolution and dose difference (DD) and distance-to-agreement (DTA) criteria impact on computation time. Conclusion will be given in Section 4.

## 2. Methods and Materials

### *2.1 A modified γ-index algorithm*

The γ-index is the minimum Euclidean distance in normalized dose-distance space (Low *et al.* 1998):

$$\gamma(\mathbf{r}_r) = \min\{\Gamma(\mathbf{r}_r, \mathbf{r}_e)\},$$

with

$$\Gamma(\mathbf{r}_r, \mathbf{r}_e) = |\tilde{\boldsymbol{r}}_r - \tilde{\boldsymbol{r}}_e|, \forall\{\mathbf{r}_e\},$$
$$\tilde{\boldsymbol{r}}_r = \left(\frac{\mathrm{r}_r}{\Delta d}, \frac{D_r(\mathrm{r}_r)}{\Delta D}\right),$$
$$\tilde{\boldsymbol{r}}_e = \left(\frac{\mathrm{r}_e}{\Delta d}, \frac{D_e(\mathrm{r}_e)}{\Delta D}\right). \tag{1}$$

Here, $D_r(\mathbf{r}_r)$ is the reference dose distribution at position $\mathbf{r}_r$ and $D_e(\mathbf{r}_e)$ is the evaluated dose distribution at position $\mathbf{r}_e$. $\Delta D$ and $\Delta d$ refer to dose DD criterion and DTA criterion, respectively. Using geometric method (Ju *et al.* 2008), the accurate Γ can be obtained by calculating the distance from the reference point $\tilde{\boldsymbol{r}}_r$ to the continuous evaluation surface formed by discrete evaluation points $\tilde{\boldsymbol{r}}_e$. And the minimum Γ value is achieved by accelerated exhaustive search with pre-sorting algorithm (Wendling *et al.* 2007). The algorithm A1 illustrates the CPU implementation of combined presorting and geometric γ-index algorithm.

**Algorithm A1:** A modified γ-index calculation algorithm implemented on CPU

---

1. Calculate the maximum DD: $\max\big(DD(\mathbf{r}_r)\big) = \max\big(D(\mathbf{r}_r) - D(\mathbf{r}_r)\big)$, $\forall\{\mathbf{r}_r\}$;
2. Calculate the geometric distance set $L_n$ which defines the maximum search range for each reference point;

   $$L_n = \sqrt{(i\Delta x)^2 + (j\Delta y)^2 + (k\Delta z)^2}/\Delta d;$$

   with $|i(or\ j,k)| \le \frac{\max(DD(\mathbf{r}_r))}{\Delta \boldsymbol{D}} \frac{\Delta \boldsymbol{d}}{\Delta \boldsymbol{x}(\boldsymbol{or}\ \Delta \boldsymbol{y}, \Delta \boldsymbol{z})}$ & $|L_n| < \frac{\max(DD(\mathbf{r}_r))}{\Delta \boldsymbol{D}}$**;**
3. Sort the geometric distance set $\{n,\ L_n\}$ in ascending order of $L_n$;
4. For each reference dose point:
   a. Set $\gamma(\mathbf{r}_r) = DD(\mathbf{r}_r)/\Delta D$;
   b. For n = 1: N (N is the length of $\{L_n\}$)
      i. Calculate Euclidean distance $\Gamma(\tilde{\boldsymbol{r}}_r,\ S_n)$ from reference point $\tilde{\boldsymbol{r}}_r$ to a $k$-simplex $S_n$: $\Gamma(\tilde{\boldsymbol{r}}_r,\ S_n) = \min\left|\tilde{\boldsymbol{r}}_r - \sum_{i=1}^{k+1} \omega_i v_i\right|$;
      ii. If $\Gamma(\tilde{\boldsymbol{r}}_r,\ S_n) > L_n$: break;
      iii. If $\Gamma(\tilde{\boldsymbol{r}}_r,\ S_n) < \gamma(\mathbf{r}_r)$ : $\gamma(\mathbf{r}_r) = \Gamma(\tilde{\boldsymbol{r}}_r,\ S_n)$;
   END

---

Similar to Wendling *et. al.* (Wendling *et al.* 2007), at the Steps 2 and 3 of Algorithm A1, we establish a sorted table of normalized geometric distance $\{L_n\}$ of all the voxels in the maximum search range. However, instead of using a manually selected search range as in (Wendling *et al.* 2007), we choose search radius $\max\big(DD(\mathbf{r}_r)\big)\,\Delta d/\Delta D$, which can avoid overestimating $\gamma$-index value.

The $\Gamma(\tilde{\boldsymbol{r}}_r\,, S_n)$ in Algorithm A1 Step 4 is obtained when $\{\omega_1, \cdots, \omega_k\} = (V^T V)^{-1} V^T P,\ \omega_{k+1} = 1 - \sum_{i=1}^{k} \omega_i$, where $P$ and $V$ are $K \times 1$ and K× $k$ matrices with a form $P = \begin{Bmatrix} c_1(\tilde{\boldsymbol{r}}_r) - c_1(v_{k+1}) \\ \vdots \\ c_n(\tilde{\boldsymbol{r}}_r) - c_n(v_{k+1}) \end{Bmatrix}, V = \begin{Bmatrix} c_1(v_1) - c_1(v_{k+1}) & \cdots & c_1(v_k) - c_1(v_{k+1}) \\ \vdots & \vdots & \vdots \\ c_n(v_1) - c_n(v_{k+1}) & \cdots & c_n(v_k) - c_n(v_{k+1}) \end{Bmatrix}$. Here, for a $k$-dimensional dose distribution $K = k + 1$, $c_j(q)$ is the $j$th coordinate of point of $q$ . Regarding $\Gamma(\tilde{\boldsymbol{r}}_r\,, S_n)$ calculation, we follow the computational acceleration techniques presented by Ju *et. al.* (Ju *et al.* 2008), where the computation is conducted recursively in the simplexes set and the recursive computation is only limited to the subset of simplexes where corresponding weights $\omega_i$ are negative. Detailed information regarding $\Gamma(\tilde{\boldsymbol{r}}_r\,, S_n)$ calculation can be found in the reference (Ju *et al.* 2008).

### *2.2 GPU implementation*

In this work, we implement the γ-index algorithm (Algorithm A1) on GPU using Compute Unified Device Architecture (CUDA) programming environment. In the Algorithm A1 Step 4, for each reference point the minimum Γ value is searched around the reference point in a search range of a radius $(DD(\mathbf{r}_r)/\Delta D)\Delta d$. On CPU, Step 4 is repeated for all reference points in a sequential manner. On GPU, this step can be parallelized for a large number of reference points and executed simultaneously using multiple threads. A key point of the GPU implementation of this algorithm is to ensure all threads in the same batch (strictly speaking *warp* in CUDA terminology) to have similar numbers of arithmetic operations. This is because, if some threads in a warp require much longer execution time, the other threads in this warp will finish first and then wait in idle until the longer execution time threads finish, implying a waste of computational power. Therefore, directly mapping the CPU version γ-index algorithm (Algorithm A1) onto GPU cannot guarantee that all threads in a warp have similar computation burden, and consequentially cannot achieve maximum speed up. As we know, the upper boundary of the search range for each reference point is $(DD(\mathbf{r}_r)/\Delta D)\Delta d$. The computation task for each reference point is then approximately proportional to the dose difference $DD(\mathbf{r}_r)$. The larger the $DD(\mathbf{r}_r)$, the more evaluation dose points will be involved, leading to longer computation time. We therefore pre-sort the voxels according to $DD(\mathbf{r}_r)$ (for convenience we call it DD sorting) and perform γ-index calculation on GPU according to pre-sorted voxel order. This DD-sorting procedure, along with pre-sorting the geometric distance set $\{n, L_n\}$, can be parallelized using recently developed Thrust library functions (Hoberock *et al.* 2010), which can sort a (or multiple) millions-element array(s) within subseconds. The completed GPU-based γ-index algorithm is illustrated as following:

**Algorithm A2:** A modified γ-index calculation algorithm implemented on GPU

---

1. Transfer dose distributions data from CPU to GPU;
2. CUDA Kernel 1: calculate in parallel the dose difference
   $$DD(\mathbf{r}_r) = D(\mathbf{r}_r) - D(\mathbf{r}_r)\,,\ \forall\{\mathbf{r}_r\};$$
3. Sort in parallel {Voxel Index, $DD(\mathbf{r}_r)$} array pair in ascending order of $DD(\mathbf{r}_r)$ using Thrust parallel sorting function and obtain $\max\big(DD(\mathbf{r}_r)\big)$;
4. CUDA Kernel 2: calculate in parallel the geometric distance set $\{L_n\}$;
5. Sort in parallel the geometric distance set $\{n, L_n\}$ in ascending order of $L_n$ using Thrust parallel sorting function;
6. CUDA Kernel 3: calculate in parallel the $\gamma$-index values using the algorithm illustrated in Step 4 of Algorithm A1 ;
7. Sort {Voxel Index, $\gamma$ } back to the original voxel index order;
8. Transfer the $\gamma$-index data from GPU to CPU.

---

We would like to point out that the Step 4-b-i of Algorithm A1 utilizes a recursive algorithm for computing $\Gamma(\tilde{\mathbf{r}}_r, S_n)$ only in the subset of simplexes which involves many IF conditions and thus creates a branching issue in GPU implementation. To avoid this problem, in Step 6 (Kernel 3) of Algorithm A2, we calculate $\Gamma(\tilde{\mathbf{r}}_r, S_n)$ in all simplexes.

## 3. Experimental Results and Discussion

### *3.1 Experimental data sets*

We tested our GPU implementation on eight IMRT dose-distribution pairs (4 lung cases (L1-L4) and 4 head-neck cases (H1-H4)), which were generated using a Monte Carlo dose engine called MCSIM (Ma *et al.* 2002) as well as an in-house developed pencil beam algorithm (Gu *et al.* 2009). Monte Carlo dose calculation results were treated as the reference dose distributions while results obtained from the pencil beam algorithm were used as the evaluation dose distributions. All the doses were originally calculated with the voxel size of $4.0\text{mm} \times 4.0\text{mm} \times 2.5\text{mm}$ and normalized to the prescription dose and interpolated to various resolution levels for comparison studies. CPU computation was conducted on a 4-core Intel Xeon 2.27 GHz processor. GPU computation was performed on one single NVIDIA Tesla C1060 card. We would like to point out that our GPU implementation did not affect the calculation accuracy; in all scenarios, the γ-index values calculated on GPU agree with those calculated on CPU within ~$10^{-6}$. In the following sections, we present results under various conditions. For the CPU implementation based on Algorithm A1, we divide the total computation time $T^C$ into two parts, *i.e.*, $T^C = T_p^C + T_\gamma^C$, where $T_p^C$ is the data processing time (Steps 1, 2, and 3 of Algorithm A1) and $T_\gamma^C$ is the γ-index calculation time (Step 4 of Algorithm A1). For the GPU implementation based on Algorithm A2, we split the total computation time $T^G$ into three parts, *i.e.*, $T^G = T_t^G + T_p^G + T_\gamma^G$, where $T_t^G$ is the data transferring time between CPU and GPU (Steps 1 and 8 of Algorithm A2), $T_p^G$ is the data processing time (Steps 2-5 and Step 7 of Algorithm A2), and $T_\gamma^G$ is the γ-index calculation time (Step 6 of Algorithm A2) .

### *3.2 Speedup of GPU vs. CPU*

For this study, we set the resolution of dose distributions to be $256 \times 256 \times 144$ (or $160, 206$) and use 3% for DD criterion and 3 mm for DTA criterion. Table 1 lists computation time for the CPU implementation (Algorithm A1) and the GPU implementation (Algorithm A2). We present two speedup factors in Table 1, with and without CPU-GPU data transferring time, *i.e.*, $T^C/T^G$ and $T^C/(T^G - T_t^G)$. These two speedup factors are quite similar (within 3% for all cases), indicating that the data transferring time in GPU calculation is not significant compared to γ-index computation time $T_\gamma^G$ . We can also see that the data processing time in both CPU and GPU implementations is relatively insignificant compared to the γ-index calculation time (Step 4 of Algorithm A1 and Step 6 of Algorithm A2). Overall, the GPU implementation can achieve about 20x~30x speedup compared to its CPU implementation.

Table 1: Calculation time of $\gamma$-index for CPU and GPU implementations for 8 IMRT dose distribution pairs.

| Case | Voxel number | CPU (sec) | | | GPU (sec) | | | | Speedup factor | |
|---|---|---|---|---|---|---|---|---|---|---|
| | | $T_p^C$ | $T_\gamma^C$ | $T^C$ | $T_t^G$ | $T_p^G$ | $T_\gamma^G$ | $T^G$ | $T^C/(T^G - T_t^G)$ | $T^C/T^G$ |
| L1 | 256×256×206 | 0.33 | 64.93 | 65.26 | 0.07 | 0.18 | 2.78 | 3.03 | 22.05 | 21.54 |
| L2 | 256×256×160 | 0.24 | 65.64 | 65.89 | 0.06 | 0.15 | 2.40 | 2.61 | 25.84 | 25.25 |
| L3 | 256×256×160 | 0.28 | 101.46 | 101.74 | 0.06 | 0.14 | 3.77 | 3.97 | 26.02 | 25.53 |
| L4 | 256×256×160 | 0.25 | 30.10 | 30.35 | 0.06 | 0.14 | 0.86 | 1.06 | 30.35 | 28.63 |
| H1 | 256×256×144 | 0.49 | 47.73 | 47.95 | 0.05 | 0.11 | 2.28 | 2.44 | 20.06 | 19.65 |
| H2 | 256×256×144 | 0.45 | 242.23 | 242.68 | 0.05 | 0.12 | 7.56 | 7.73 | 31.60 | 31.39 |
| H3 | 256×256×144 | 0.24 | 116.14 | 116.38 | 0.05 | 0.12 | 4.76 | 4.93 | 23.85 | 23.61 |
| H4 | 256×256×144 | 0.22 | 107.61 | 107.86 | 0.05 | 0.12 | 4.21 | 4.38 | 24.91 | 24.63 |

### *3.3 The effect of DD sorting on computation time $T_\gamma^G$*

As mentioned in Section 2.2, introducing of DD sorting (Step 3 in Algorithm A2) can better synchronize the computational tasks on CUDA threads and consequently to reduce the computation time. We illustrate the effect of DD sorting on $T_\gamma^G$ in Table 2. The speedup achieved by DD sorting is around 2.3-4.9 times.

Table 2: Speedup achieved in GPU computation by sorting voxels based on the dose difference values.

| Case | $T_\gamma^G$ (Non-DD sorting) (sec) | $T_\gamma^G$ (DD sorting) (sec) | Speedup factor achieved by DD sorting |
|---|---|---|---|
| L1 | 7.37 | 2.78 | 2.65 |
| L2 | 7.47 | 2.40 | 3.11 |
| L3 | 11.00 | 3.77 | 2.91 |
| L4 | 1.99 | 0.86 | 2.31 |
| H1 | 8.48 | 2.28 | 3.72 |
| H2 | 24.50 | 7.56 | 3.24 |
| H3 | 15.00 | 4.76 | 3.15 |
| H4 | 20.6 | 4.21 | 4.90 |

### *3.4 The effect of $\gamma$-index value on computation time $T_\gamma^C$ and $T_\gamma^G$*

From Table 1, we see that, both $T_\gamma^C$ and $T_\gamma^G$ change significantly from case to case. Take cases L3 and L4 as an example, they have the same number of voxels, but their $\gamma$-index calculation time differs by more than 3 times. We know that computation time $t$ for each reference point is proportional to the number of voxels searched $N_s$, *i.e.*, $t \propto N_s$. The relationship of $N_s$ with the search length $L$ can be expressed as $N_s \propto L^n$, where $n$ is the dimension of dose distributions. Here, for all the testing cases in this paper, $n = 3$. On the other hand, from Algorithm A1 Step 4-b-ii, we can deduce the search length $L$ is proportional to the $\gamma$ value at each reference point, *i.e.*, $L \propto \gamma$. Thus, we can state that the computation time $t$ for each reference point is proportional to $\gamma^n$: $t \propto \gamma^n$. In Figure 1(a), we plot $T_\gamma^C$ and $T_\gamma^G$ versus the summation of $\gamma^3$ over all voxels ($\sum \gamma^3$) for each of 8 testing cases, respectively. We see that both $T_\gamma^C$ and $T_\gamma^G$ are monotonically increasing with the value of $\sum \gamma^3$. To further illustrate our point, we choose one set of patient data (H2) and shift the evaluation dose distribution (normalized to the prescription dose ) by -10%, -9%, ..., up to 10%, at a step size of 1%, inside the region of 10% iso-dose line. Figure 1(b) illustrates the $\gamma$-index calculation time $T_\gamma^C$ and $T_\gamma^G$ with respect to $\sum \gamma^3$. We can see that $T_\gamma^C$ versus $\sum \gamma^3$ can be fitted with a straight line (dashed line in Figure 1(b)), indicating that $T_\gamma^C$ strictly follows the rule $T_\gamma^C \propto \sum \gamma^n$. However, the date points for $T_\gamma^G$ are much more scattered. This is because the GPU computation time is not only the function of $\sum \gamma^3$, but also the function of $\gamma^n$ distribution that determines the variation of threads computation time in a warp.

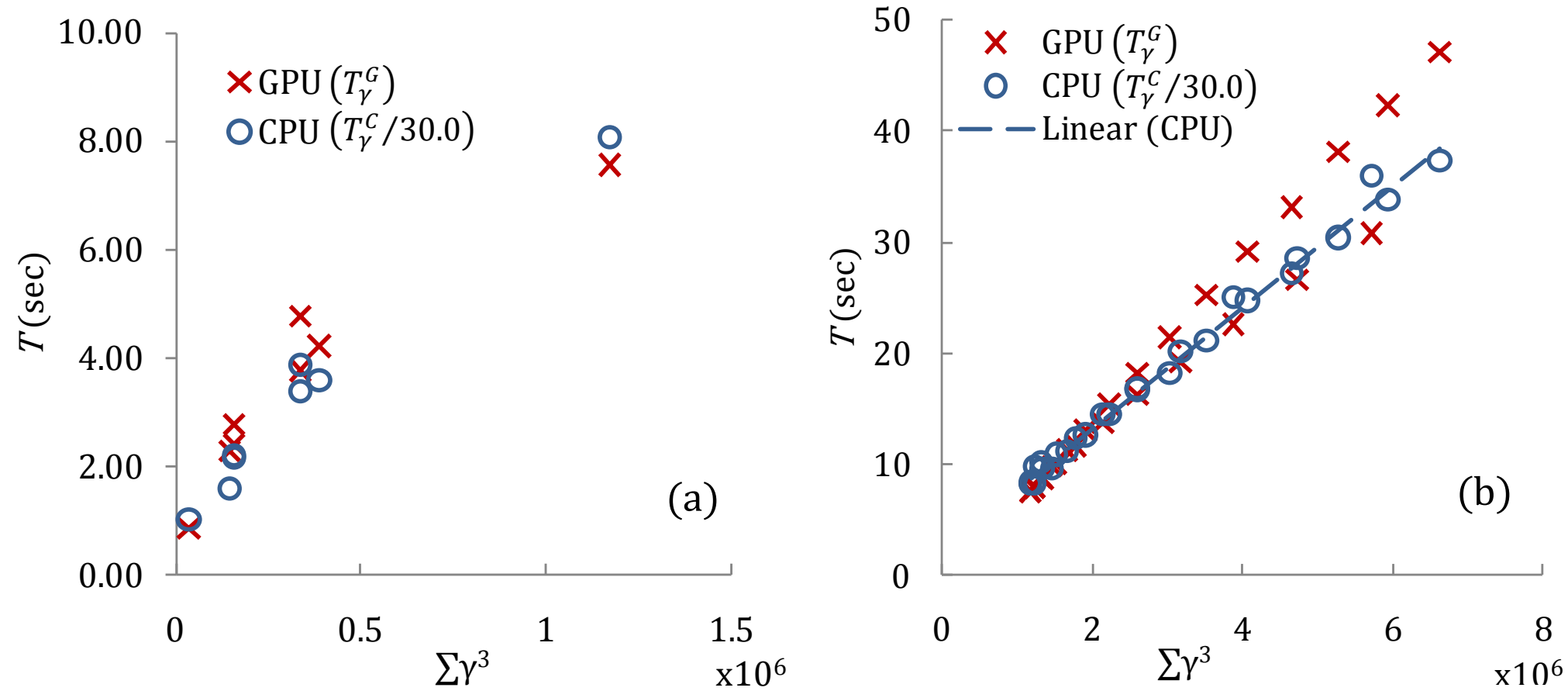

Figure 1: (a) GPU and CPU computation time for eight testing cases vs. the summed $\gamma^3$ value. (b) GPU and CPU computation time for case H2 with various dose shifts on evaluation dose distribution vs. the summed $\gamma^3$ value. For convenient purpose, we scaled down CPU computation time by a factor of 30.0 to illustrate them in the same vertical axis of GPU computation time.

### *3.5 The effect of dose distribution resolution on the computation time $T_\gamma^C$ and $T_\gamma^G$*

We choose the case H2 to test the effect of the dose distribution resolution on computation time $T_\gamma^C$ and $T_\gamma^G$. We interpolate the dose distributions to various resolution levels, including $128 \times 128 \times 144$, $256 \times 256 \times 72$, $256 \times 256 \times 144$, and $512 \times 512 \times 72$. We illustrate $T_\gamma^C$ and $T_\gamma^G$ changes with respect to the resolution changes in Figure 2. As indicated by the power trend lines (dashed lines in Figure 2), $T_\gamma^C$ increases approximately as $N^{1.90}$ while $T_\gamma^G$ increases approximately as $N^{1.68}$, when the resolution of dose distribution increases $N$ times. As illustrated in Algorithm A1, the CPU based $\gamma$-index calculation is completed with two loops. The outer loop is over all the reference dose points and the inner loop is the exhaustive search in a limited region around each reference dose point. The computation time of the outer loop is increased linearly with respect to the increase of resolution of dose distributions. The inner loop computation time is proportional to the number of voxels involved. For the geometric method, the number of involved voxels in a fixed region is increased linearly as the resolution increases. Overall, it leads to a quadratic increase of computational time ($\propto N^{1.90}$) for a linear change of resolution. For the GPU algorithm A2, the computation time increases as $(N)^{1.68}$ in this testing case. This slight difference might be due to the fact that the memory accessing time can be hidden by large arithmetic operations in GPU computation.

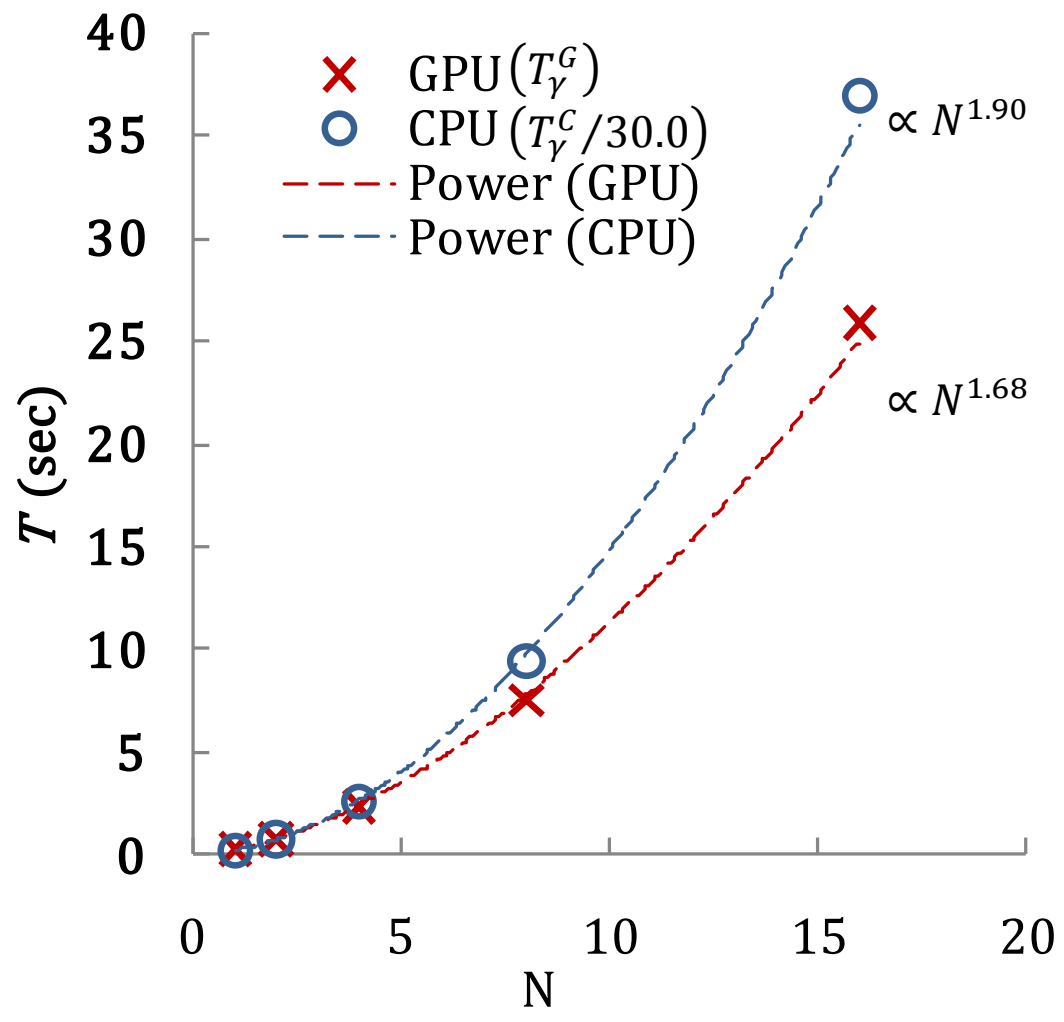

Figure 2. CPU and GPU computation time as functions of dose distribution resolution. Again, For convenient purpose, we scaled down CPU computation time by a factor of 30.0 to illustrate CPU computation time in the same axis of GPU computation time.

*3.6 The effect of DD and DTA criteria on computation time $T_\gamma^C$ and $T_\gamma^G$*

In this study, we choose case H2 and fix the resolution to $256 \times 256 \times 144$, then vary DD and DTA criteria. Table 3 lists the computation time obtained from varying criteria. There are three interesting phenomena: 1) when we increase DD criterion value and fix the DTA criterion value, the computation time decreases; 2) when we fix DD criterion value and increase DTA criterion value, the computation time decrease; 3) when we increase both DD criterion value and DTA criterion value proportionally, for example, from 1%, 1mm to 2%, 2mm, or 3%, 3mm, the computation time does not change. As we mentioned in Section 3.4, the $\gamma$-index calculation time for each reference dose point $t \propto \gamma^n$. For phenomenon 1), when we increase DD criterion value and fix DTA criterion value, the $\gamma$-index value decreases. Consequently, the required searching steps decreases, and computation time decreases. For phenomenon 2), when we fixe DD criterion value, but increase DTA criterion value by $k$ times, the $\gamma$-index values will decrease by $k'$ times, with $k' < k$, which decreases computation time by $(k')^n$ times. However, when DTA criterion value increases by $k$ times, the resolution in the normalized dose-distance space will also increase by $k^n$ times, which consequently increases the computation time by $k^n$ times. The net change of the computation time should be $(k/k')^n$. Since $k' < k$, the overall computation time will then increase. For phenomenon 3), when we increase both DTA and DD criteria values simultaneously by $k$ times, the $\gamma$-index values decrease by $k' = k$ times. The increase rate of computation time will be $(k/k)^n = 1$. The computation time under this situation will not change. From Table 3, we can see that the change of DD and DTA criteria values does not affect the speedup factor achieved with GPU implementation.

Table 3: CPU and GPU computation time varies with DD and DTA criteria values.

| DD criteria | DTA criteria (mm) | Computational time (sec) | | Speedup factor |
|---|---|---|---|---|
| | | $T_\gamma^G$ | $T_\gamma^C$ | $T_\gamma^C/T_\gamma^G$ |
| 1% | 1 | 7.56 | 240.37 | 31.79 |
| 1% | 2 | 26.31 | 863.33 | 32.81 |
| 1% | 3 | 52.96 | 1545.09 | 29.17 |
| 2% | 1 | 2.27 | 74.27 | 32.72 |
| 2% | 2 | 7.56 | 265.82 | 35.16 |
| 2% | 3 | 15.62 | 486.68 | 31.16 |
| 3% | 1 | 1.27 | 40.09 | 31.57 |
| 3% | 2 | 3.70 | 119.32 | 32.25 |
| 3% | 3 | 7.56 | 245.34 | 32.45 |

## 4. Conclusions

In this paper, we implemented a modified γ-index algorithm on GPU. We evaluated our GPU implementation on eight pairs of IMRT dose distributions. Overall, our GPU implementation has achieved about 20x~30x speedup compared to the CPU implementation and can finish the $\gamma$-index calculation within a few seconds. We also studied the effects of various factors on the calculation time on both CPU and GPU. We found that the pre-sorting procedure based on the dose difference speeds up the GPU calculation by about 2~4 times. The CPU computation time is proportional to the summation of $\gamma^n$ over all voxels, where $n$ is the dimension of dose distributions. The GPU computation time is approximately proportional to the summation of $\gamma^n$ over all voxels, but affected by the variation of $\gamma^n$ among different voxels. We also found that increasing the resolution of dose distribution leads to a quadratic increase of computation time on CPU, while less-than-quadratic increase on GPU. We observed that both CPU and GPU computation time decrease when increasing DD criterion value and fixing DTA criterion value, increase when increasing DTA criterion value and fixing DD criterion value, and don't vary when DD criterion value and DTA criterion value both change proportionally. Both CPU and GPU codes developed in this work for γ-index dose evaluation are in public domain and available upon request.

**Acknowledgements**

This work is supported in part by the University of California Lab Fees Research Program and by an NIH/NCI grant 1F32 CA154045-01. We would like to thank NVIDIA for providing GPU cards for this project.